\documentclass{aastex}
\usepackage{emulateapj5}
\usepackage{graphicx}
\usepackage{times}
\newcommand{\gl}{$\tilde{g}$}
\newcommand{\OSO}{${\rm S}^0$~}
\newcommand{\SO}{{\rm S}^0}
\newcommand{\fmc}{fm$^{-3}\;$}
\newcommand{\MeVcs}{$\mbox{MeV}\mbox{c}^{-2}\;$}
\newcommand{\gmcmc}{$\mbox{gm}\mbox{cm}^{-3}\;$}

\slugcomment{NYU-TH-00/09/05}
\makeatletter

\newenvironment{inlinefigure}{%
\def\@captype{figure}%
\noindent\begin{minipage}{0.999\linewidth}\begin{center}}
{\end{center}\end{minipage}\smallskip}

\begin{document}
\title{Neutron Stars with a Stable, Light Supersymmetric Baryon}

\author{Shmuel Balberg\altaffilmark{1}, Glennys~R.~Farrar\altaffilmark{2} and 
Tsvi Piran\altaffilmark{3}}

\altaffiltext{1}{Department of Physics, 
University of Illinois at Urbana--Champaign, 1110 West Green Street,
Urbana, IL 61801--3080, sbalberg@astro.physics.uiuc.edu}
\altaffiltext{2}{Department of Physics, New York University, NY, NY 10003,
farrar@physics.nyu.edu}
\altaffiltext{3}{Racah Institute of Physics, 
The Hebrew University, Jerusalem 91904, Israel, tsvi@nikki.phys.huji.ac.il}

%%%%%%%%%%%%%%%%%%%%%%%%%%%%%%%%%%%%%%%%%%%%%%%%%%%%%%%%%%%%%%%%%%%%%%%%%%
\begin{abstract}
If a light gluino exists, the lightest gluino-containing
baryon, the \OSO, is a possible candidate for 
self-interacting dark matter.  In this scenario, the simplest
explanation for the observed ratio $\Omega_{dm}/\Omega_b \approx 6-10$
is that $m_{S^0} \sim 900\;$\MeVcs; this is not at present excluded by
particle physics. Such an \OSO could be present in neutron 
stars, with hyperon formation serving as an intermediate stage. We calculate
equilibrium compositions and equation of state for high density matter with
the \OSO, and find that for a wide range of parameters the properties of 
neutron stars with the \OSO are consistent with observations. In particular, 
the maximum mass of a nonrotating star is $1.7-1.8\;M_\odot$, and the presence 
of the \OSO is helpful in reconciling observed cooling rates with hyperon 
formation.
\end{abstract}
%%%%%%%%%%%%%%%%%%%%%%%%%%%%%%%%%%%%%%%%%%%%%%%%%%%%%%%%%%%%%%%%%%%%%%%%%%%


\section{Introduction}

The interiors of neutron stars offer a unique meeting point between
astrophysics on one hand and nuclear and particle physics on the other. 
The macroscopic properties of neutron stars, such as mass,
radius, spin rate and thermal history depend on the
microscopic nature of matter at very high densities. 
Neutrons stars are, in this sense,
cosmological laboratories of physics at densities exceeding the
nuclear saturation density ($\rho_0\simeq0.16\;$\fmc).
One aspect of matter at supernuclear densities is the possible appearance of
hadronic degrees of freedom beyond nucleons. The principle candidates
are hyperons, meson condensates and a deconfined quark phase
(see, e.g., Glendenning 1996, Prakash et al.~1997 for reviews). 

In this work we wish to illuminate another possibility, the appearance of an 
$R$-baryon. We are motivated by the observation of \citet{SS_SIDM} that the 
observed properties of dark matter (DM) on scales of order 10-100 kpc are 
better described by a theory with self-interacting DM than by conventional, 
weakly interacting DM. Since the required cross-section to mass ratio, 
$\sigma_{XX}/m_X\sim 1$b/GeV, is characteristic of hadronic physics at low 
energies, a light stable neutral hadron would be a welcome feature of 
physics-beyond-the standard model, and such particles arise naturally in 
theories with a very light gluino.  In some sense, 
this is the minimal extension of the standard model needed to produce the
required DM particle, and therefore merits examination.

A supersymmetric version of the standard model contains a massless
gluino (\gl), the spin-1/2 superpartner of the gluon;  therefore it is
endowed with a corresponding chiral symmetry called $R$-invariance.
When supersymmetry is broken, the $R$-invariance may or may not break. If it
does not, the gluino mass arises only from loops sensitive to the
Higgs vevs (which do break $R$-invariance) and is naturally of order
$100\;$\MeVcs. Although considerable effort has been devoted to excluding
it, a very light gluino remains a possibility -- mainly because of the
difficult of making firm, first-principles predictions for
non-perturbative phenomena in QCD. The question deserves very serious
scrutiny because the existence of a light gluino would resolve several
puzzles. These issues and the constraints imposed by experiment are
reviewed by Farrar (1998, 2000). See also \citet{fg00ltgl} and 
references therein.

The lightest supersymmetric ``$R$-hadrons'' are the spin-1/2 \gl$g$ bound 
state ($R^0$ or glueballino) and the spin-0 baryon $uds$\gl~ bound state, 
the \OSO. The \OSO is thus a boson with baryon number unity 
and a strangeness charge $-1$. The strong color hyperfine interaction 
unique to the $uds$\gl~in a flavor singlet state makes the mass of the \OSO 
much less than that of all other $R$-baryons \citep{Farrar84,BuccFarr85}, and 
probably less than 2 GeV$c^{-2}$, which would make it stable through 
conservation of $R-$parity \citep{Farrar96}.  

In the very early universe the \OSO remains in equilibrium with
nucleons via processes such as $R^0 N \leftrightarrow \SO K$, down
to $T \approx 35$ MeV (Farrar, Spergel \& Steinhardt 2000; see also in 
Wandelt et al.~2000). Hence  for a suitable \OSO mass, 
slightly less than the proton mass, standard freezeout estimates lead to a 
density ratio $\Omega_{\SO}/\Omega_B \approx 6-10$, as required if the \OSO 
is to be the dark matter.  To posit a new baryon lighter than a
nucleon is quite shocking!  But 
it is known that the color hyperfine force is very strong in this
channel, and there are innumerable examples of the inadequacy of
phenomenological models of non-perturbative QCD phenomena. The
possibility of such a light \OSO should not be dismissed on
theoretical grounds, and we must seek observational constraints. 
One possible astrophysical constraint is whether this new baryon can form in 
neutron stars, and if so, would it imply unacceptable neutron star properties? 

Nucleons cannot convert directly into the \OSO, even if the latter is 
less massive. Conservation of $R-$parity requires that $R$-hadrons are 
produced in pairs, forbidding direct conversion of individual nucleons into 
\OSO particles.  The weak neutral current process $nn\rightarrow{\rm
S}^0{\rm S}^0$ requires simultaneous creation of two units of
strangeness ($\Delta S=-2$).  Strangeness-changing neutral currents
($\Delta S = \pm 1, \Delta Q = 0$) are highly suppressed in the standard 
model due to the GIM mechanism. Thus, $nn\rightarrow{\rm S}^0{\rm S}^0$ is 
doubly GIM suppressed; 
crude estimates indicate that its rate is of order the inverse Hubble time 
\citep{FSScoming}. However processes such as 
$\Lambda\Lambda\rightarrow{\rm S}^0{\rm S}^0$ ($\Delta S=0$) and even 
$\Sigma^-p \rightarrow {\rm S}^0{\rm S}^0$ (charged current, $\Delta S=-1$) 
are not GIM suppressed, so that if the conditions in interiors of neutron 
stars allow for hyperons to appear (as is predicted by many works on this 
issue, see, e.g., Glendenning 1985, Weber \& Weigel 1989, Knorren, 
Prakash \& Ellis 1995, Schaffner \& Mishustin 1996, Balberg \& Gal 1997, 
Huber et al.~1998, Baldo, Burgio \& Schulze 2000), these hyperons
could serve as a doorway for \OSO formation, which might affect observable 
properties of neutron stars.  

Our goal in this {\it letter} is to explore whether a light \OSO can form in 
neutron stars and assess whether the resulting global properties of these 
stars are compatible with observations. In \S~\ref{sect:interactions} we
extend an equation of state for high density matter with hyperons to
include the \OSO.  The equilibrium composition with the \OSO is found
in \S~\ref{sect:composition}, followed by a discussion of the
corresponding properties of neutron stars in \S~\ref{sect:EoSandNS}.

\section{\OSO interactions in high density matter}
\label{sect:interactions}

In order to include \OSO particles in an equation of state of high density
matter, we must specify their interactions with other baryons and among
themselves. Here we do so in the framework of an effective equation of state
(Balberg \& Gal 1997, Balberg, Lichtenstadt \& Cook 1999), where the 
potential felt by a single, isospin singlet particle of species $x$ placed in 
matter composed of species $y$ at a baryon number density $\rho_y$ is
\begin{equation}\label{eq:effpot}
V_x(\rho_y)=
    a_{xy}\rho_y+c_{xy}\rho_y^{\gamma}+w\rho_y^{\theta}\;.
\end{equation}
In equation~(\ref{eq:effpot}) $a$ is negative and represents a long range,
attractive component in the interaction, $c$ is positive, corresponding to an
intermediate
range repulsive component, and $w$ is also positive, representing a
universal repulsive short range interaction, presumably arising from
the quark structure of the baryons. The powers $\gamma$ and $\theta$ must be
greater than unity, so that repulsion dominates at high densities.

For nucleons and hyperons, the values of the coefficients can be
constrained by experimental data. Since no such data
exists for the \OSO, we resort to phenomenological arguments to
estimate the magnitude of these terms. {\it The attractive} component 
% of the interaction for baryons 
is generally associated with two-pion or
two-kaon exchanges. This interaction is strong because the
intermediate state in the $s$-channel can be the same as the initial
state, hence the amplitude has a small energy denominator.  However
the flavor singlet
\OSO cannot exchange $\pi$ or $K$ without becoming a much more massive
flavor octet $R$-baryon \citep{BuccFarr85}.  Thus the long-range
attractive interaction is strongly suppressed for the \OSO. A more
empirical way to arrive at the same conclusion is to note that if
there were a significant long-range attractive force the \OSO would
bind to nuclei (and to itself), and nuclei with exotic mass values have not 
been observed.  We therefore assume that the attractive component
is negligible for the \OSO$\!-$related interactions. 
{\it The intermediate repulsive} component may be associated with the
exchange of vector mesons, such as the $\rho$ and $\eta$. Only the
flavor-singlet component of these vector-mesons can couple to the
\OSO, so we roughly assume that the magnitude of this term (the value
of $c$ in Eq.~[\ref{eq:effpot}]) is half of its value for the
$\Lambda$.  {\it The universal short-range} interaction for the \OSO
is taken as identical to that of the baryons. Here we use EoS 1 of
\citet{BLCRoles}, where $\gamma=\frac{4}{3}$ and
$\theta=\frac{5}{3}$, so that for the \OSO $a=0$ and
$c=329.5\;\mbox{MeV}\mbox{fm}^4$ for all its interactions, and
$w=223.6\;\mbox{MeV}\mbox{fm}^5$.  Below we refer to this choice of
parameters along with an \OSO mass of $m_{{\rm S}_0}=900\;$\MeVcs as
our nominal set.  Most of our conclusions prove insensitive to these
choices, as will be seen below.

\section{Composition of neutron star matter with the \OSO}
\label{sect:composition}

The composition of neutron star matter is
determined by chemical equilibrium for all weak-interaction processes.
Specifically, these are $B_1\rightarrow B_2  \ell \nu_{\ell}$, where
$B_1$ and $B_2$ are baryons, $\ell$ is a lepton and $\nu_{\ell}$ is its
corresponding neutrino (or antineutrino).
Neutron star matter is well approximated as cold ($T=0$) and neutrino free
($\mu_\nu=0$) so that all equilibrium conditions are summarized concisely
in the form:
\vspace{-0.25cm}
\begin{equation}\label{eq:mueq}
\mu_{B_i}=\mu_n-q_{B_i} \mu_e\;,
\vspace{-0.25cm}
\end{equation}
where $\mu_{B_i}$ is the chemical potential of the baryon species
$B_i$, $\mu_n$ and $\mu_e$ are the neutron and electron chemical
potentials, respectively, and $q_{B_i}$ is the electric charge of the
baryon $B_i$.  The equilibrium composition at any baryon number
density, $\rho_B$, is then determined by two additional conditions:
$\Sigma_i x_{B_i}=1$ and charge neutrality, $\Sigma_i
x_{B_i}q_{B_i}+\Sigma_i x_{\ell} q_{\ell}=0$, where $x_{B_i}$
and $x_{\ell}$ denote the fraction per baryon for each baryon and
lepton species, respectively.

Equation~(\ref{eq:mueq}) does not fully apply to the case of the \OSO,
since the process $nn\rightarrow {\rm S}^0{\rm S}^0$
is suppressed, even if energetically favorable. The \OSO must
maintain a direct equilibrium only with hyperons, i.e.,
$ 2\mu_{{\rm S}^0}=2\mu_{Y^0}$, and $2\mu_{{\rm S}^0}=\mu_{Y^-}+\mu_p$,
but $2\mu_{{\rm S}^0}=2\mu_n$ is {\it not} required 
(where $Y^0$ and $Y^-$ denote a neutral and a negatively charged hyperon,
respectively). The significance of this distinction is that a combination 
of \OSO and nucleons may coexist even while $\mu_{{\rm S}^0}<\mu_n$, as long 
as no hyperons are present. On the other hand, a finite hyperon fraction will 
indirectly impose that $\mu_{{\rm S}^0}=\mu_n$, so then 
equation~(\ref{eq:mueq}) applies to all species.

We emphasize that even if $\mu_{{\rm S}^0}\leq\mu_n$, nuclear matter
cannot ``burn'' exothermally into \OSO matter. Unlike the case of
``strange-quark-stars'' \citep{Witten84,AFO86}, where a
deconfining front converts an entire neutron star into the (hypothetically)
lower energy strange quark matter if a seed of of strange
quark matter is somehow formed \citep{Olinto87}, there exists no equivalent
stimulant to initiate conversion of nuclear-matter into \OSO-matter,
regardless of density or composition.

In Fig.~\ref{fig:comp}a we show for
reference the equilibrium composition of neutron star matter as a
function of baryon density, using the ``standard'' equilibrium
composition of matter with nucleons and hyperons but no \OSO (hereafter, 
NH matter) based on EoS 1 of \citet{BLCRoles}, assuming that
the equation of state is valid up to densities as large as $10\rho_0$.
The results, with hyperon formation beginning at about $2\rho_0$, are
typical of matter with hyperons 
\citep{GlenH85,WWH89,KnorrenalH95,SMH96,BGH97,HuberalH98,BaldoalH00}.

Figure \ref{fig:comp}b depicts the equilibrium composition when the \OSO
is included according to the assumptions of \S~\ref{sect:interactions}.
Due to the low mass of the \OSO, its chemical potential is lower than that
of all hyperons, even when the fractions of the latter are infinitesimally
small. As a result, all hyperons that form convert to \OSO
particles, and the matter consists only of nucleons and
\OSO. The composition is determined by the nucleons being in
equilibrium with an exactly zero fraction of hyperons
\footnote{\small{In reality, 
some additional nucleons would convert to hyperons through quantum and thermal 
($T \sim 0.01$MeV in neutron star cores)
fluctuations, and these hyperons would decay to \OSO's. The actual 
equilibrium composition would be one where the chemical potential of the 
nucleons is slightly below that of zero-fraction hyperons at every density 
(in other words, fluctuations would deplete a thin outer layer in phase space 
of nucleons). Assuming that the probability of fluctuations decreases 
exponentially with the energy gap, the quantitative 
effect on the results should be small.}}. Even though absent
from the matter, the hyperons do play a role in determining the
amount of \OSO particles which are created.

%%%%%%%%%%%%%%%%%%%%%%%%%%%%%%%%%%%%%%%%%%%%%%%%%%%%%%%%%%%%%%%%%%%%%%%
\begin{inlinefigure}
\centerline{\includegraphics[width=1.0\linewidth]{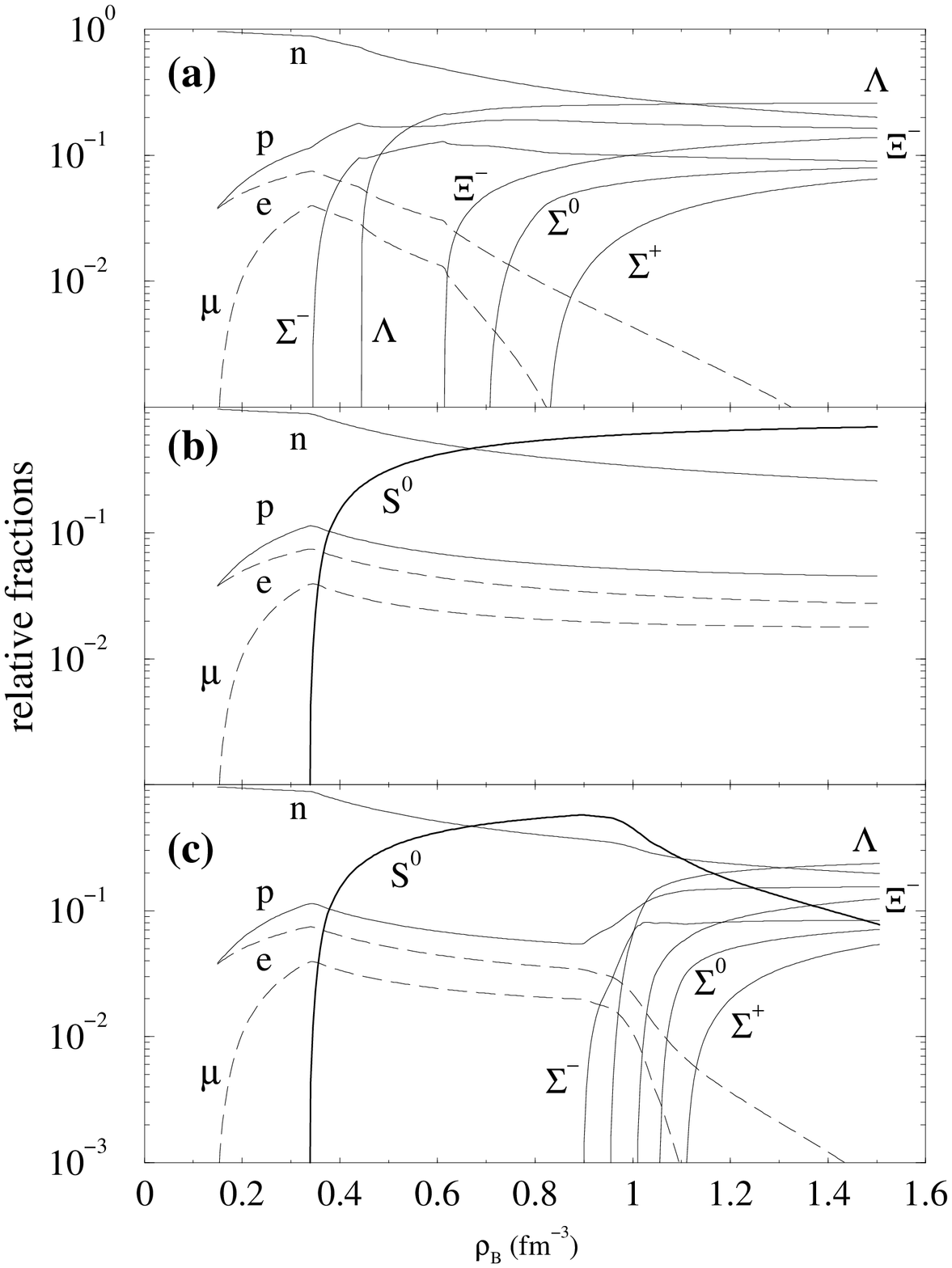}}
\figcaption{Equilibrium compositions for high density matter for EoS 1 of 
ref.~[1]: (a) matter with hyperons when the \OSO is ignored,
(b) including the \OSO with $m_{{\rm S}^0}=900\;$\MeVcs, and
(c) same as (b) but with $m_{{\rm S}^0}=975\;$\MeVcs. \label{fig:comp}}
\end{inlinefigure}
%%%%%%%%%%%%%%%%%%%%%%%%%%%%%%%%%%%%%%%%%%%%%%%%%%%%%%%%%%%%%%%%%%%%%%%%

Since the \OSO is never in chemical equilibrium with the nucleons
(or the zero fraction hyperons), the equilibrium composition would be
identical for any $m_{{\rm S}_0}<900\;$\MeVcs. In fact,  
for the nominal interactions, the same composition is also found for 
$m_{{\rm S}_0}\lesssim m_p$. A qualitative change is found when 
$m_{{\rm S}_0}$ is larger: in Fig.~\ref{fig:comp}c we show the equilibrium 
composition for nominal \OSO interactions, but 
with $m_{{\rm S}^0}\!=\!975\;$\MeVcs.
At low and intermediate densities the composition is identical to that of
the $m_{{\rm S}^0}=900\;$\MeVcs case, but beginning at a density of
0.9 \fmc the larger mass of the \OSO makes it favorable for
a finite fraction of hyperons to coexist in the matter. The general
equilibrium condition (Eq.~[\ref{eq:mueq}]) is now satisfied for
all particle species. Owing to the \OSO having only repulsive
interactions, the hyperons achieve chemical equilibrium
with it even though it has a lower mass and no kinetic
energy. Moreover, for larger densities, it becomes increasingly
favorable to reduce the \OSO fraction as the equilibrium composition
gradually approaches that of NH matter.
With similar calculations we find that for nominal interactions,
if $m_{{\rm S}^0}\simeq1030\;$\MeVcs the \OSO appears only after some
hyperons already exist in the matter, and if
$m_{{\rm S}^0}\gtrsim1060\;$\MeVcs \OSO formation is suppressed 
%altogether
for the entire range of densities.

\section{Equation of State and Mass Sequences}\label{sect:EoSandNS}

We now turn to examine the equation of state
(pressure vs.~mass-energy density, hereafter EoS) of high density matter with 
the \OSO. These are shown in Fig.~\ref{fig:EoS}, including the models used for
Fig.~\ref{fig:comp} and also a model labeled N for nuclear matter, 
%%%%%%%%%%%%%%%%%%%%%%%%%%%%%%%%%%%%%%%%%%%%%%%%%%%%%%%%%%%%%%%%%%%%%%%%
\begin{inlinefigure}
\centerline{\includegraphics[width=1.0\linewidth]{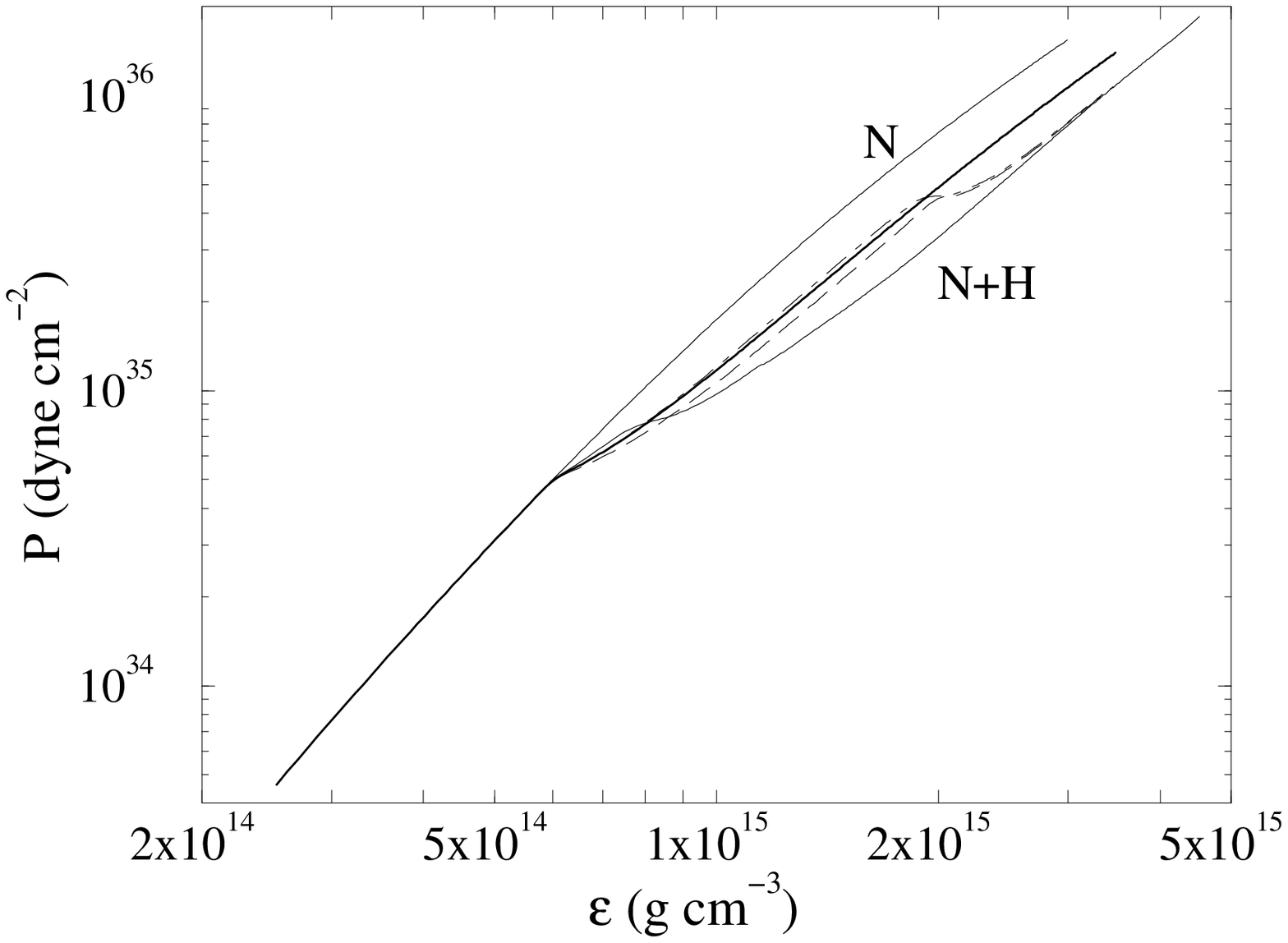}}
\figcaption{Equation of state for matter of the compositions shown
in Fig.~1. Matter with the \OSO: nominal values (see text) -
thick solid line; nominal interactions and $m_{{\rm S}^0}=975\;$\MeVcs -
dashed line; nominal mass and increased repulsion - dot-dashed line.
Also shown are the equations of state for nucleon and hyperon matter - N+H,
and nuclear matter - N. \label{fig:EoS}}
\end{inlinefigure}
%%%%%%%%%%%%%%%%%%%%%%%%%%%%%%%%%%%%%%%%%%%%%%%%%%%%%%%%%%%%%%%%%%%%%%%%
where 
hyperon (and therefore also \OSO) formation is ignored.

The hierarchy of EoSs in Fig.~\ref{fig:EoS} is mostly
determined by the number of available species in each model.
Nuclear matter naturally has the stiffest EoS, followed
by matter with the \OSO. Even though the \OSO, being a boson, does
not exert any Fermi pressure, the fact that four species of hyperons are
suppressed due to its appearance makes such matter stiffer than NH matter,
which is clearly the softest.  An exception occurs just above the
threshold density for hyperon formation, where NH matter includes only a
single hyperon species (the $\Sigma^-$), so that the contribution of its
Fermi pressure does make NH matter stiffer.

Although the mass of the \OSO does not affect its contribution to the
pressure, it does contribute to the mass energy-density. Hence a
lower $m_{{\rm S}^0}$ yields a stiffer EoS, while a larger
value yields a softer one. For $m_{{\rm S}^0}\!=\!975\;$\MeVcs
such softening is enhanced at a density of $\sim1.8\times 10^{15}\;$\gmcmc,
when hyperons begin to accumulate; as the \OSO fraction is depleted
this EoS gradually approaches that of NH matter.
It is noteworthy that increasing the repulsive power of the \OSO does not
lead to significant stiffening of the EoS. We demonstrate
this with another model shown in Fig.~\ref{fig:EoS} when
$m_{{\rm S}^0}\!=\!900\;$\MeVcs but 
$c=494.5\;\mbox{MeV}\;\mbox{fm}^4$ (50\%
larger than the nominal value). Magnified repulsion leads to an increase of
the \OSO chemical potential, which in turn enhances hyperon formation.
The resulting composition of this model is quite similar to that of
Fig.~\ref{fig:comp}c, and the EoS, while slightly stiffer than
nominal until hyperons appear, is then softened. We emphasize that this is
a manifestation of the ability of hyperons to exert ``pressure control''
\citep{BGH97,BLCRoles}, since the extent of their formation tracks the
dependence of particle interactions on density.

Given an EoS, the structure of a spherical, non-rotating neutron star is 
found by numerically integrating the Tolman-Oppenheimer-Volkoff equations of 
general relativistic hydrostatic equilibrium (see, e.g., 
Misner, Thorne \& Wheeler 1973). The resulting mass sequences
(stellar mass as a function of the assumed central density) for the EoSs
of Fig.~\ref{fig:EoS} are given in Fig.~\ref{fig:Mvsrho}.
All sequences are identical up to about $1.1\;M_\odot$, where the mass
sequences diverge according to the hierarchy of EoSs.
The maximum mass found for matter with the \OSO is $1.81\;M_\odot$
for the nominal case, $1.68\;M_\odot$ for $m_{{\rm S}^0}=975\;$\MeVcs,
%%%%%%%%%%%%%%%%%%%%%%%%%%%%%%%%%%%%%%%%%%%%%%%%%%%%%%%%%%%%%%%%%%%%%%%%
\begin{inlinefigure}
\centerline{\includegraphics[width=1.0\linewidth]{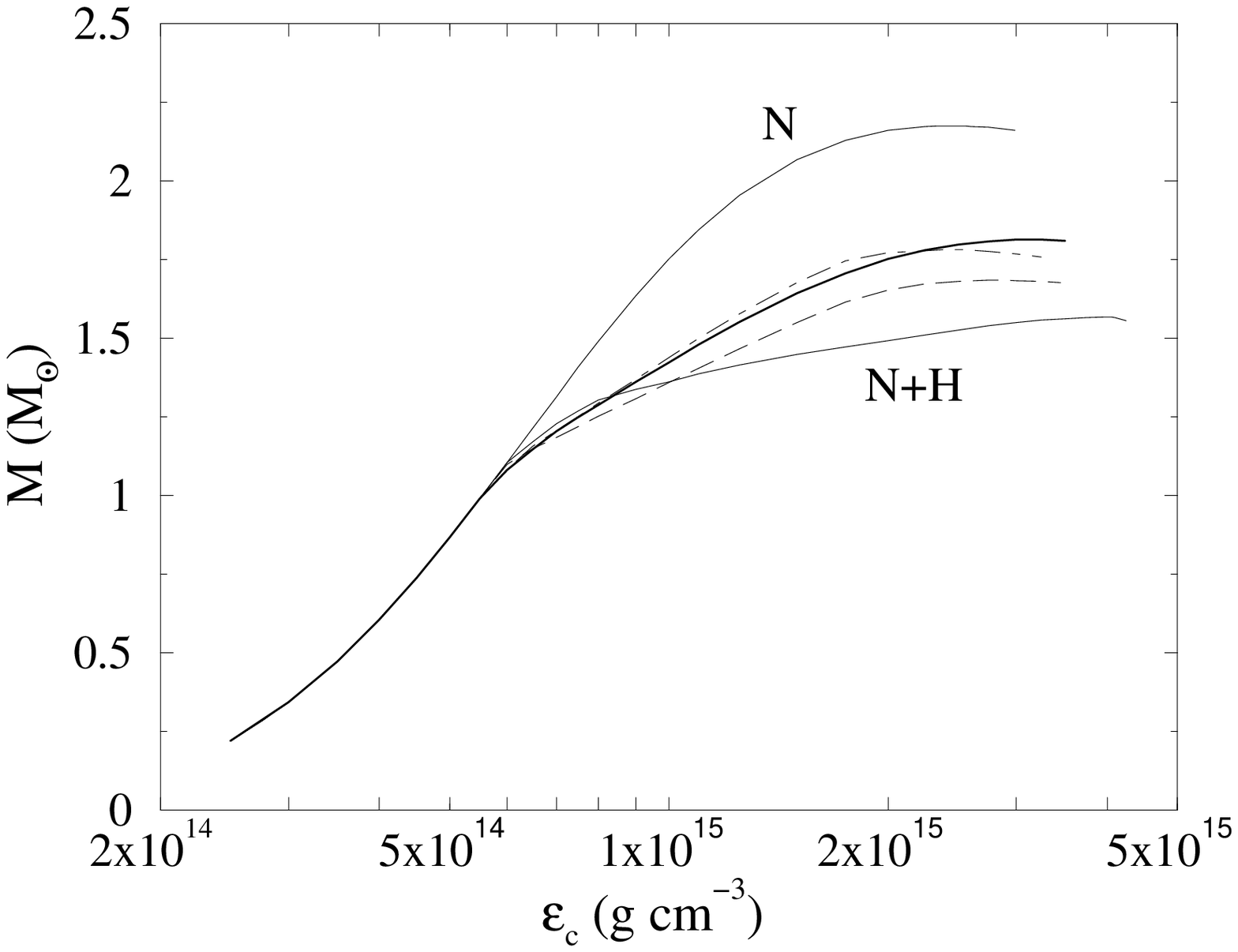}}
\figcaption{Nonrotating neutron star sequences for the five equations of state
shown in Fig.~2. \label{fig:Mvsrho}}
\end{inlinefigure}
%%%%%%%%%%%%%%%%%%%%%%%%%%%%%%%%%%%%%%%%%%%%%%%%%%%%%%%%%%%%%%%%%%%%%%%%
\nolinebreak
and $1.78\;M_\odot$ for $m_{{\rm S}^0}=900\;$\MeVcs but enhanced \OSO
repulsive interactions. In both latter cases, hyperons coexist with the \OSO
only in the largest mass stars, those of mass closer than $0.1\;M_\odot$ to
the maximum mass in each case. We note that for EoS 1,
the maximum masses for nuclear matter and NH matter are $\sim2.18$ and
$\sim1.57\;M_\odot$ respectively.

At present, the primary observational constraint on high density
equations of state is that the maximum mass must be at least 
$1.44\;M_\odot$, the inferred mass of PSR 1913+16.
All EoSs discussed above, with and without the \OSO, are
acceptable.  This limit does allow to rule out very soft EoSs, 
such as would arise if the \OSO had a low mass {\it and} an
attractive component in its interactions. Such interactions would make
a pair of \OSO similar to the hypothetical H dibaryon - a six quark
flavor singlet, boson (Jaffe 1977a, 1977b) - whose interactions are supposed 
to be similar to those of two $\Lambda$ hyperons combined. The EoS for
matter with the H would indeed be too soft to sustain a
$1.44\;M_\odot$ star, unless its mass is significantly above
$2m_\Lambda$ \citep{GSBtheH}. The expected strongly repulsive
interactions of the \OSO are thus crucial for allowing its existence in
neutron stars.

\section{Conclusions and Discussion}\label{sect:concdis}

Our primary conclusion is that the existence of a light ($m_{S^0}\lesssim
m_p$) \OSO particle is consistent with the fundamental observed
properties of neutron stars, namely, their inferred masses.  This
conclusion applies for a wide range of models for high density matter
with the \OSO.  In particular, the maximum mass of a static neutron
star under nominal assumptions regarding the \OSO is found to be about
$1.8\;M_\odot$. We note that since it suppresses hyperon accumulation,
the appearance of the \OSO yields a stiffer equation of state and
higher maximum mass than obtained with nucleons and hyperons without
\OSO.  Therefore, identification of a relatively high mass neutron star 
would be easier to reconcile with an \OSO
component than in its absence (where hyperons accumulate
undisturbed). Some indication for the existence of such massive neutron stars 
is implied by recent analyses, although uncertainties are still
significant \citep{vKW}. We emphasize that {\it current} data on
neutron star masses is not restrictive enough to decisively 
constrain the properties of high density matter.

The presence of \OSO particles could have a significant and welcome
effect on the cooling rate of neutron stars, if they do indeed replace
hyperons in the cores. Hyperons can participate in the direct Urca
cooling process ($B_1\rightarrow B_2 e\nu$) while the \OSO cannot. The
threshold fraction required for hyperons to drive direct Urca cooling
processes is of the order of 0.01 \citep{PPLP92}, so neutron stars with
hyperons are susceptible to rapid cooling, which seems inconsistent
with observations \citep{Schaabal96}.  Hyperon direct Urca cooling can
be partially suppressed if the $\Lambda$ hyperons pair to a $^1{\rm
S}_0$ superfluid state \citep{SBSB98}, but the extent of suppression is
sensitive to the details of the equilibrium composition and gap energy
\citep{Pageal00}. On the other hand, \OSO formation would prevent the
hyperon direct Urca, or restrict it only to the highest mass stars, so
that very moderate cooling rates are attainable. Future measurements
of surface thermal emission from young neutron stars with the {\it
CHANDRA} and {\it XMM} satellites may lead to stronger
constraints on cooling rates; if so this will yield valuable insight 
into the internal structure of high density matter.

%\acknowledgements

The research of GRF is supported in part by NSF-PHY-99-96173.
The idea that the \OSO could be a dark matter candidate, and
constraints on such a scenario, were developed by GRF in collaboration
with D.~Spergel and P.~Steinhardt; that work will be reported in
greater depth elsewhere \citep{FSScoming}; a brief description can be found 
in \citep{wandelt00}. SB and TP
thank NYU for hospitality which facilitated this research.

%\newpage

\end{document}